\documentclass[12pt]{article}                            
\usepackage{psfrag}
\usepackage{graphicx}                                                                                
\usepackage{rotating}                                                                                
\usepackage{longtable}                                                                                
\usepackage{supertabular}                                                                                
\usepackage{amssymb}                                                                                
\usepackage{amsmath}

\title{Effect of an external magnetic field on the nematic-isotropic phase 
transition in mesogenic systems of uniaxial and biaxial molecules\\[8mm]
} 
\author{Nababrata Ghoshal, Kisor Mukhopadhyay$^1$ and
Soumen Kumar Roy$^{2,}$
\footnote{Corresponding author. E-mail: skroy@phys.jdvu.ac.in,
Tel: +91 9874741525; fax: +91 33 24146584}\\
Department of Physics, Mahishadal Raj College,\\[1mm]
Mahishadal, Purba-Medinipur, West Bengal, India\\[1mm]
$^1$Department of Physics, Sundarban Mahavidyalaya,\\[1mm]
Kakdwip, South 24 Parganas, West Bengal, India\\[1mm]
$^2$Department of Physics, Jadavpur University,\\[1mm]
Kolkata - 700 032, India}
\date{  }

\begin{document}

\maketitle
\begin{abstract}
\noindent Influence of an external magnetic field on the 
nematic-isotropic ($N-I$) 
phase transition in a dispersion model of nematic liquid crystals, 
where the molecules are either perfectly uniaxial or biaxial (board-like),
has been studied by Monte Carlo simulation. Using multiple histogram 
reweighting technique and finite size scaling analysis the order of the 
phase transition, the transition temperature at the thermodynamic limit 
and the stability limit of the isotropic phase below 
the transition temperature for different 
magnetic field strengths have been determined. 
The magnetic field dependence of the shift in $N-I$ 
transition temperature is observed to be more rapid than that
 predicted by the standard Landau-de Gennes and 
Maier-Saupe mean field theories. 
We have shown that for a given field 
strength the shift in the transition temperature is higher for the biaxial molecules 
in comparison with the uniaxial case. The study shows that the $N-I$ transition 
for the biaxial molecules is weaker than the well known weak first order $N-I$ transition
for the uniaxial molecules and the presence of the external magnetic 
field (up to a certain critical value) makes the transition much 
more weaker for both the systems.
The estimate of the critical magnetic field ($\sim 110~T$) 
for the common 
nematics is found to be smaller than the earlier estimates.
\end{abstract}

\section{INTRODUCTION}
The study of the behaviour of liquid crystals in presence of an external field 
(either electric or magnetic) has long been an active area of research
because of its fundamental and technological importance. 
There have been several theoretical \cite{fre,han,woj,fan,she,hor,rem} 
and experimental \cite{hel,nic,lel1,lel2,rosen,osta} reports 
dealing with different phase transition phenomena in presence of 
external electric and magnetic fields. 

In absence of any external field nematic liquid crystals composed of 
anisotropic molecules, usually modeled as rod-like in shape, possess 
long-range orientational order due to sufficiently strong intermolecular  
forces and in the uniaxial nematic phase ($N$) the molecules become aligned, 
on the average, 
along a single macroscopic direction called the director ($\boldsymbol{n}$). 
The degree of this orientational order is characterized by the 
second rank tensor nematic order parameter $\boldsymbol{Q}$ \cite{degen}.  
With increasing temperature the order parameter of a thermotropic
liquid crystal in the nematic phase decreases  
and jumps to zero as the liquid crystal undergoes a weakly first-order 
nematic-isotropic 
($N-I$) transition at a temperature $T_{NI}$.
The orientational order no longer exists in the isotropic phase ($I$).  
If an external field is applied, the director in the $N$ phase, 
tends to align parallel to the direction of the field provided the 
liquid crystal material has a positive dielectric or diamagnetic anisotropy 
and becomes pinned at relatively weak fields although the orientational 
order increases only by a small amount. 
The transition temperature $T_{NI}$ also 
gets shifted towards a higher value.
In the presence of an external field a weak orientational 
order is induced even in the isotropic phase resulting in an 
 anisotropic phase known as the $\textit{paranematic}$ phase ($pN$). 
From theoretical studies \cite{han,woj} it has been 
observed that when the external field
is weak the $N-I$ (or $pN$) phase transition occurs with a 
finite jump in the nematic order parameter and the transition is first order in nature. 
As the field strength is increased the difference between the $N$ phase
and the $pN$ phase decreases and finally it vanishes at a classical 
critical point defined by a critical temperature ($T_C$) and a critical 
field ($B_C$). This critical end point is analogous to the classical critical 
points observed in liquid-vapour and ferromagnetic systems \cite{stanley}. 
Above the critical point the $I$ (or $pN$) and the $N$ phases become
indistinguishable, the symmetries of the phases being the same.  
The effect of an external magnetic field on the $N-I$ transition
is analogous to the effect observed by the application of pressure 
at the liquid-gas transition, the paranematic phase being the analog of the 
liquid-gas coexistence phase.  

There are several reports \cite{hel,lel1,lel2,dhara} of experimentally observed 
electric field induced first-order isotropic-nematic phase transition since its 
earliest theoretical prediction based on the mean field theories 
of Maier-Saupe and Landau de-Gennes \cite{degen}.
On the other hand similar studies in the presence of a magnetic field are relatively fewer.
The influence of a magnetic field on the $N-I$ transition was first studied by 
Rosenblatt \cite{rosen}. His experimental observation demonstrated  
that the presence of a magnetic field of strength ${14.8}$~$T$
shifts the $N-I$ transition temperature only by a few milliKelvin.
The reason behind this small change is the very low value of 
the anisotropy of 
diamagnetic susceptibility for the traditional liquid crystal 
materials \cite{degen}.  
In order to substantially alter the $N-I$ transition temperature for 
the conventional calamitic liquid crystals the necessary magnetic 
field strength should be larger than $100$~$T$
and the estimated critical value of the magnetic field is 
much higher than this \cite{rosen,shen}.  

The experimental observation of a magnetic field induced 
first-order isotropic-nematic transition in a thermotropic liquid crystal 
has been reported only recently by Ostapenko et al \cite{osta} who
also observed a much higher shift in the $N-I$ transition temperature 
($\sim{0.8}^\circ C$ at $B=23$~T).
Observation of such pronounced effects  
became possible by the use of a new class (non-traditional)
of liquid crystal molecules 
having a bent core  and by using a high-field resistive magnet. 
However, Ostapenko et. al. could not reach the critical point of the 
system in their investigation where a magnetic field up to
$31$~T was used.

The bent-core compounds have exhibited different fascinating phenomena 
in liquid crystal science, for example, the formation of the long searched 
biaxial nematic phase in thermotropic liquid crystals \cite{madsen,achar}
and the formation of 
chiral phases resulting from achiral molecules \cite{link}.
Another experimental study \cite{wiant} has shown that the $N-I$ 
transition in the bent-core compounds is more weakly first order 
as the value of $T_{NI}-T_- \approx 0.4^\circ~C$ which is 
significantly below the values $\sim 1^\circ~C$ in typical calamitics,
$T_-$ being the super cooling limit of the isotropic phase. 
The exceptional behaviour of 
this new class of liquid crystals has been explained 
\cite{wiant} by considering 
the isotropic phase to be composed of microscopic complexes or "clusters"
of bent core molecules.     
Another important aspect of the bent-core 
mesogenic molecules is their inherent biaxiality. 
A mean-field study \cite{rem} was reported more than two 
decades ago in which 
the effect of an external magnetic field on biaxial nematogenic molecules was 
examined. It is evident from this work that as the degree of biaxiality is 
increased the critical field strength drops rapidly and  
for a given field strength the shift in $T_{NI}$ is 
higher in uniaxial nematics composed of biaxial molecules.
So far as our knowledge is concerned no other theoretical 
or computer simulation 
study on systems composed of biaxial molecules coupled with 
external magnetic field has been reported. 

In this paper we have tried to investigate using Monte Carlo (MC) simulation
the effect of an external magnetic field on the $N-I$ (or $N-pN$)
transition. Using a lattice model of nematics where the particles interact
via a dispersion potential we have performed the multiple histogram 
reweighting \cite{ferr} and finite-size scaling analysis \cite{lee1,lee2}
of our data. We have studied the magnetic field dependence of the transition
temperature $T_{NI}$ for systems composed of either uniaxial or 
biaxial molecules 
and have compared our results with that predicted by the mean field 
theory. We have also confirmed the first orderedness of the $N-pN$ 
phase transition and find that the transition gets weaker with the 
increase in the magnetic field and ultimately reaches a critical point.
Using the technique similar to that applied by Saito et. al. in a 
recent work \cite{saito} for the study of quark mass dependence of 
the finite temperature QCD phase transition, we have estimated the value
of the critical magnetic field for the system composed of uniaxial 
molecules. 
The use of finite size scaling along with the multiple histogram reweighting
has also enabled us to obtain the magnetic field dependence of the 
supercooling limit of the isotropic phase (or $pN$ phase ) in the thermodynamic 
limit. Expectedly we find that the difference $T_{NI}(B)-T_-(B)$ 
decreases with the increase in the magnetic field and vanishes at a certain
point ($B_C$, $T_C$) which we have identified as the critical point.
Our work is intended to supplement the predictions of the 
mean field theory in this area where no other theoretical work has 
been reported. The MC simulations which have so far been performed
to study the effects of external fields on the $N-I$ transitions
are the works of Luckhurst and coworkers \cite{luck1} and that of Berardi 
et. al. \cite{berar}. However these simulations do not address the features 
of the phase transitions we have described.

In the next section we describe the model employed in our simulation.
This is followed by the computational details, results and conclusions.

\section{THE DISPERSION POTENTIAL AND THE EFFECT OF THE MAGNETIC FIELD}
We consider a system of biaxial prolate molecules
possessing $D_{2h}$ symmetry (board-like), whose centres of mass are
associated with a simple-cubic lattice
and subjected to an external magnetic field.
The total energy of the 
system is the sum of two terms: (i) a dispersion potential term which
takes into account the interaction between all nearest neighbour pairs 
of molecules and (ii) a field term which represents the interaction 
of each molecule with the external field.    
The dispersion term in the Hamiltonian contains a factor $\lambda$
which is a measure of the molecular biaxiality. The case of uniaxial 
molecules ($D_{\infty h}$ symmetry) is obtained by simply setting  this 
parameter equal to zero.

The orientationally anisotropic dispersion pair interaction obtained 
from London dispersion model \cite{buck,stone} explicitly 
depends on both the mutual orientation of the two interacting molecules 
(say the $i^{th}$ and $j^{th}$ molecules), 
and on their orientations with respect to the  
intermolecular unit vector ($\boldsymbol{r}_{ij}$).    
By isotropically averaging over the intermolecular 
unit vector, $\boldsymbol{r}_{ij}$ the dispersion potential between two 
identical neighbouring molecules becomes \cite{luck,bis} 
\begin{equation}\label{e1} 
U_{ij}^{disp}=-\epsilon_{ij} \{R_{00}^2(\Omega_{ij})+2\lambda[R_{02}^2(\Omega_{ij})+R_{20}^2(\Omega_{ij})]+4\lambda^2R_{22}^2(\Omega_{ij})\}.
\end{equation}
Here $\Omega_{ij}=\{\phi_{ij},\theta_{ij},\psi_{ij}\}$ denotes the triplet 
of Euler angles defining the relative orientation of $i^{th}$ 
and $j^{th}$ molecules; we have used the convention used by Rose \cite{rose}
 in defining the Euler angles. 
$\epsilon_{ij}$ is the strength parameter which is assumed 
to be a positive constant ($\epsilon$) when the particles $i$ and $j$ are 
nearest neighbours and zero otherwise. $R_{mn}^L$ are combinations of 
symmetry-adapted ($D_{2h}$) Wigner functions 
\begin{equation}\label{e2}
R_{00}^2=\frac{3}{2}\cos^2\theta-\frac{1}{2}
\end{equation}  
\begin{equation}\label{e3}
R_{02}^2=\frac{\sqrt{6}}{4}\sin^2\theta\cos2\psi
\end{equation}  
\begin{equation}\label{e4}
R_{20}^2=\frac{\sqrt{6}}{4}\sin^2\theta\cos2\phi
\end{equation}  
\begin{equation}\label{e5}
R_{22}^2=\frac{1}{4}(1+\cos^2\theta)\cos2\phi\cos2\psi-\frac{1}{2}\cos\theta\sin2\phi\sin2\psi.
\end{equation}  

The parameter $\lambda$ is a measure of the molecular biaxiality and it depends
on the molecular properties. For the dispersion interactions, $\lambda$ can be 
expressed in terms of the eigenvalues ($\rho_1$, $\rho_2$, $\rho_3$) of the 
polarizability tensor {\boldmath $\rho$} of the biaxial molecule
\begin{equation}\label{e6}
\lambda=\sqrt{\frac{3}{2}}\frac{\rho_2-\rho_1}{2\rho_3-\rho_2-\rho_1}.
\end{equation}
The condition for the maximum biaxiality is 
$\rho_3-\rho_2=\rho_2-\rho_1>0$, $\lambda=\lambda_C=1/\sqrt{6}$ 
and this self-dual geometry corresponds to the Landau point in the 
phase diagram where a direct biaxial nematic to isotropic 
phase transition occurs. $\lambda < \lambda_C$ corresponds 
to the case of prolate molecules whereas $\lambda > \lambda_C$ corresponds
to oblate molecules. 
This dispersion model can successfully reproduce both the uniaxial and the biaxial
orientational orders and various order-disorder transitions as a function of 
temperature and molecular biaxiality.
The phase diagram for this  dispersion model has been well studied 
by mean field theory as well as Monte Carlo methods \cite{luck,bis}.  
In our simulations we consider two cases - in one case $\lambda=0$ and the 
pair potential takes the usual Lebwohl-Lasher (LL) form \cite{leb} for 
nematic liquid crystals of perfectly uniaxial molecules which has been 
extensively studied by Zhang et al \cite{zhang}. In the other case we choose 
$\lambda=0.2$ which represents a biaxial system composed of prolate biaxial 
molecules. For the LL model (uniaxial one) 
there is a single weak first-order $N-I$ transition at a 
dimensionless temperature $T=1.1232\pm 0.0001$ \cite{zhang,fab}. From 
the Monte Carlo results, as reported in \cite{bis,ghoshal}, the biaxial model 
($\lambda=0.2$) exhibits a biaxial-uniaxial phase transition at low 
temperature ($T\approx 0.2$) and a uniaxial-isotropic transition at 
higher temperature ($T\approx 1.1$). The biaxial nematic-uniaxial nematic
transition is known to be second order while the uniaxial nematic-isotropic
transition is first order. 
The dimensionless temperature has been defined 
as $T=k_BT_K/\epsilon$, $T_K$ being the temperature measured in Kelvin.

The interaction of a uniform external magnetic field $\boldsymbol {B}$
chosen along the laboratory $Z$ axis (unit vector $\boldsymbol{z}$), 
with the $i$th molecule resulting from its coupling with 
the longest molecular symmetry axis $\boldsymbol{w}_i$ is taken as 
\begin{equation}\label{e7} 
U_{i}^{field}=-\epsilon\xi[\frac{3}{2}(\boldsymbol{w}_i\cdot\boldsymbol{z})^2-\frac{1}{2}].
\end{equation}
Where, $\xi$ is a dimensionless quantity which determines 
the strength of coupling of the molecular symmetry 
axis with the magnetic field and is given by 
\begin{displaymath}
\xi= \frac{(\Delta\kappa) B^2}{3\mu_0\epsilon}.
\end{displaymath}
Here, $\Delta\kappa=\kappa_\parallel-\kappa_\perp$ is the anisotropy
of the molecular magnetic polarizability and 
$\mu_0$ is the permeability of the free space.  
In the simulations we take $\xi$ to be a positive quantity so that 
the molecules tend to get their long axes aligned along the magnetic field.

The total energy $\mathcal{E}$ of the system is therefore given by
\begin{equation}\label{e8} 
\mathcal{E}=\sum_{\langle i,j\rangle}U_{ij}^{disp}+\sum_i U_i^{field}
\end{equation}
where, the $\langle ~\rangle$ bracket represents the nearest neighbours 
and the two terms on the right hand side are obtained from 
Eqs. \ref{e1} and \ref{e7}.

\section{COMPUTATIONAL ASPECTS}
To calculate the thermodynamic observables of interest as a function of 
$\lambda$, $\xi$ and $T$ we have performed a 
series of Monte Carlo (MC) simulations using the conventional 
Metropolis algorithm
on a periodically repeated simple cubic lattice,  
 for four system sizes.
A Monte Carlo move was attempted by selecting a site at random and 
then by choosing one of the laboratory axes at random and rotating the 
molecule at that site about the chosen laboratory axis following 
the Barker-Watts method \cite{bar}.   
For generating histograms of energy and the constant energy
averages of the order parameter and its square,
simulations were run in cascade, in order of increasing 
temperature $T$ for a given set of values of $\lambda$ and $\xi$. 

We have used $10^6$ sweeps or MCS (Monte Carlo steps per site)
for the equilibration and $(4-6)\times10^6$ MCS for the production 
run for every set of values of $\lambda, \xi$ and $T$.
For the largest lattice size ($L=30$), the total run length is more 
than $10~000$ times the correlation time. We have divided the total 
run into several ($100$) blocks by performing independent simulations 
for each set of values of $L$, $\lambda$, $\xi$ and $T$ so 
that we could compute the jackknife errors \cite{new}. 
In order to analyze the orientational order we have calculated the 
second rank order
parameters $\langle R_{mn}^2\rangle$ following the procedure described
by Vieillard-Baron \cite{vie}. According to this, a $\boldsymbol{Q}$ 
tensor is defined for the molecular axes associated with a 
reference molecule. For an arbitrary unit vector {\boldmath $w$}, 
the elements of the $\boldsymbol{Q}$ tensor are defined as 
\begin{equation}\label{e11}
Q_{\alpha\beta}(\boldsymbol{w}) = \langle(3 w_\alpha w_\beta-\delta_{\alpha\beta})/2\rangle
\end{equation}
where the average is taken over the configurations and the subscripts $\alpha$ 
and $\beta$ label Cartesian components of {\boldmath $w$} 
with respective to an 
arbitrary laboratory frame. By diagonalizing the matrix one obtains nine 
eigenvalues and nine eigenvectors which are then recombined to give 
the four order parameters 
$\langle R_{00}^2\rangle$, $\langle R_{02}^2\rangle$, 
$\langle R_{20}^2\rangle$ and $\langle R_{22}^2\rangle$  with respect to 
the director frame \cite{camp}. 

Out of these four second rank order parameters 
the usual uniaxial order parameter  $\langle R_{00}^2\rangle$ (or, $S$)
which measures the alignment of the longest molecular symmetry axis 
with the primary director ($\boldsymbol{n}$),
is involved in our study. The full set of order parameters 
are required to describe completely the biaxial nematic phase of a 
system of biaxial molecules. In our work we have simulated 
a very short temperature range in the higher sides of $T$ 
within which no biaxial phase occurs.

We have calculated the reduced specific heat per particle and the ordering 
susceptibility 
from fluctuations in the energy and the order parameter respectively
\begin{equation}\label{e12}
C_v = \frac{\langle {E}^2 \rangle-{\langle E \rangle}^2}{Nk_B{T}^2}
\end{equation}
\begin{equation}\label{e13}
\chi = \frac{N(\langle {R_{00}^2}^2 \rangle-{\langle R_{00}^2 \rangle}^2)}{T}
\end{equation}
where $E$ is the scaled total energy of the system.    


Various thermodynamic quantities have been computed using the
multiple-histogram reweighting method proposed by Ferrenberg and Swendsen \cite{ferr}.
In our simulations we have generated 
the energy histogram for certain values of the 
dimensionless temperature $T$ and corresponding to each energy bin we 
have generated (constant energy) averages of the order parameter and 
its square. These averages are used to evaluate the 
order parameter and the corresponding
susceptibility as a function of temperature using the reweighting method.

\section{RESULTS AND DISCUSSION}
\subsection{Probability distribution functions and free energy for the uniaxial and the biaxial models}
In order to analyse the effect of the external field on the 
phase behaviour of nematic liquid
crystals we need to simulate the models 
within a very narrow range of temperature ($T=1.110$ to $1.140$) 
around $T_{NI}$. 
We aimed at getting the transition temperatures with an accuracy of 
$\pm 0.0001$ and have therefore used the multiple histogram 
reweighting technique for the purpose.

We have performed extensive 
Monte Carlo simulation for different values of the external 
field parameter $\xi$ at five or six different temperatures within the 
said temperature range to generate histograms for both the uniaxial 
and the biaxial models for $L=18, 22, 26$ and $30$. 
For the uniaxial molecules simulations 
have been performed for five different values of $\xi$ 
ranging from $0$ to $0.00125$ with an increment of $0.000375$ 
for each lattice size, while for the biaxial model five different 
values of $\xi$ from $0$ to $0.001$ with an increment 
of $0.00025$ have been used.
Thus a total of about $200$ simulations were performed for 
different values of $\lambda$, $L$, $\xi$ and $T$.
The histograms were used to obtain 
the temperature dependence of internal energy $\langle E \rangle$, 
order parameter $\langle R_{00}^2 \rangle$ and 
the corresponding response functions i.e. the specific heat ($C_v$) and 
the order parameter susceptibility ($\chi$).    

\begin{figure}[tbh]
\begin{center}
\resizebox{120mm}{!}{\rotatebox{0}{\includegraphics[scale=1.0]{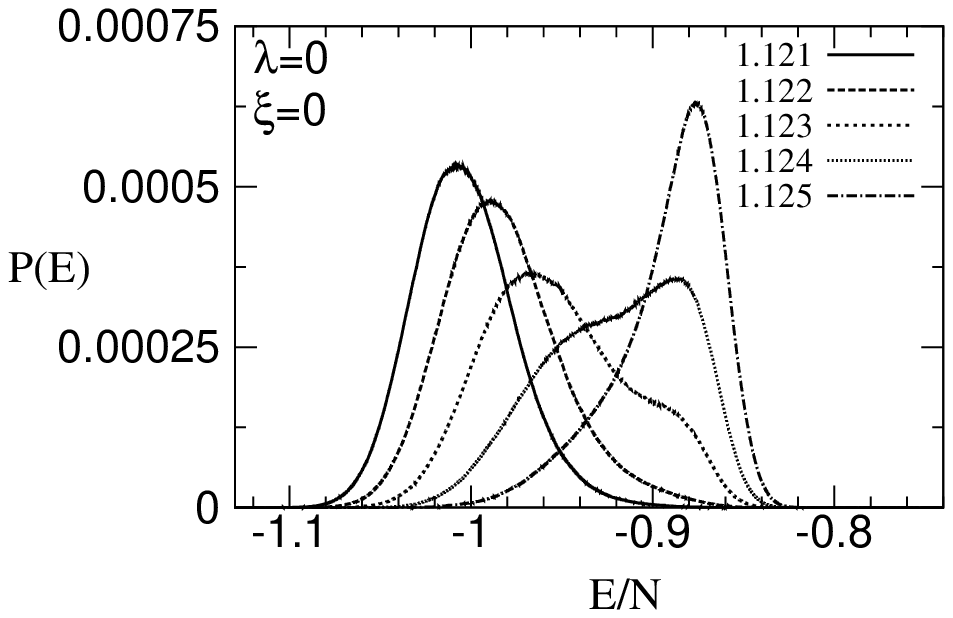}}}
\end{center}
\caption{Normalized histogram for five different temperatures  $T =$
$1.121$, $1.122$, $1.123$, $1.124$ and $1.125$ for the uniaxial model ($\lambda=0$ and $L=30$) in absence of an external magnetic field.}\label{f1}
\end{figure}

\begin{figure}[tbh]
\begin{center}
\resizebox{120mm}{!}{\rotatebox{0}{\includegraphics[scale=1.0]{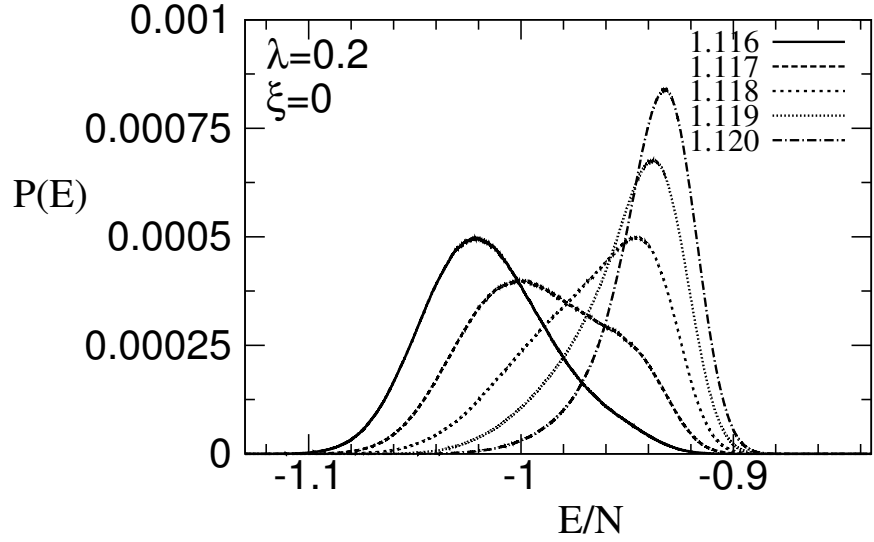}}}
\end{center}
\caption{Normalized histogram for five temperatures  
for the biaxial model ($\lambda=0.2$ and $L=30$) in absence of an external magnetic field.}\label{f2}
\end{figure}

The results for the normalized histogram count  
$P(E)=h(E)/\sum_E h(E)$ in absence of an external magnetic 
field for both $\lambda=0$ and $\lambda=0.2$ have been plotted 
in Figs. \ref{f1} and  \ref{f2} respectively for the $L=30$ lattice. 
These figures show an evidence of double peak like structures which 
are however not well separated and are more so in Fig \ref{f2}.
These merely confirm the well known weakly first order nature 
of the $N-I$ transition 
which gets weaker for the biaxial molecules. 
With increase in the strength of the magnetic field the overlap in the 
double peak like structures are found to increase even more for both
values of $\lambda$. 

\begin{figure}[tbh]
\begin{center}
\resizebox{120mm}{!}{\rotatebox{0}{\includegraphics[scale=1.0]{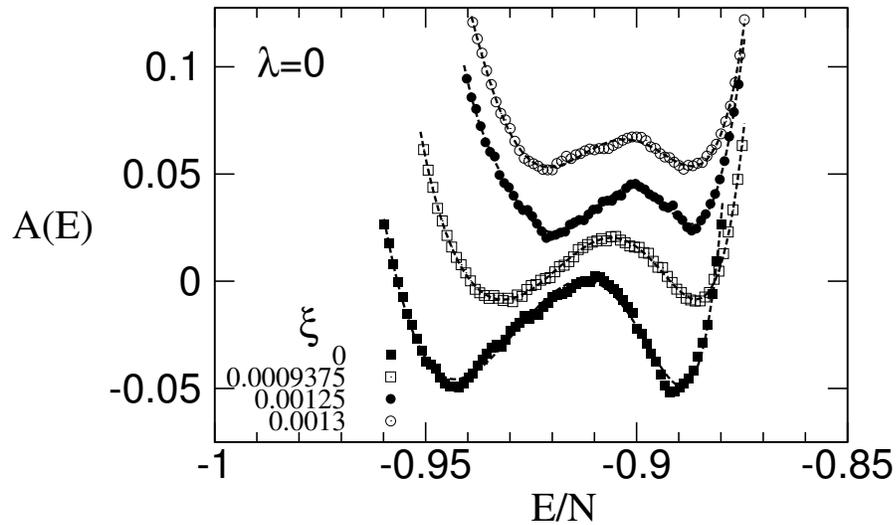}}}
\end{center}
\caption{Free energy $A$ as a function of energy per molecule 
with and without external magnetic field for the uniaxial model for $L=30$.
From bottom to top the field parameters are $0$($\blacksquare$), 
$0.0009375$($\square$), $0.00125$($\bullet$) and $0.0013$($\circ$).
An eighth-order polynomial fit to the data is also presented.}\label{f3}
\end{figure}

We have derived the relevant part of the free-enegy like functions 
$A(E)$ from the energy distribution functions, $P(E)$
for both $\lambda=0$ and $\lambda=0.2$ using the relation $A(E)=-lnP(E)$. 
For the uniaxial molecules we have shown (Fig. \ref{f3})  
the field dependence of the free energy $A$ at finite size transition 
for the largest simulated system size (i.e. for $L=30$) for which
the free energy barrier between the two minima is well pronounced and the 
effect of the external field on $\Delta F$ is clearly visible. 
Here, the bulk free energy, $\Delta F$ is given by 
$\Delta F(\xi,L)=A(E_m;T,\xi,L)-A(E_1;T,\xi,L)$ where $E_1$ is the 
energy at which either of the two minima of $A$ (of equal depth)
appears and $E_m$ gives the position of the maximum of the 
free energy $A$. 
The temperatures in all cases were adjusted to obtain two minima of 
equal depth. We observe that the double-well becomes shallower as 
$\xi$ increases. The external field thus reduces the strength of the 
first order transition and
this suggests that for a particular 
value of the field ($\xi_{c}(L)$) the barrier between the two minima 
of $A(E)$ is likely to vanish which would correspond to the end point 
where the first order phase transition turns into a crossover \cite{saito}.

\begin{figure}[tbh]
\begin{center}
\resizebox{120mm}{!}{\rotatebox{0}{\includegraphics[scale=1.0]{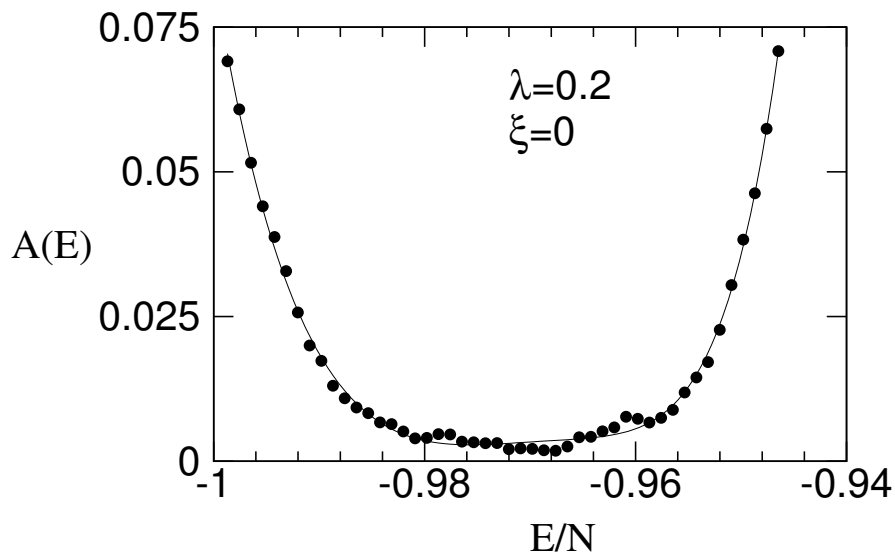}}}
\end{center}
\caption{Free energy $A$ as a function of energy per molecule 
without any external field for the biaxial model for $L=30$.}\label{f4}
\end{figure}

The mesogenic molecules having biaxiality $\lambda=0.2$ however
does not show any noticeable energy barrier 
separating the two minima in $A$, 
 even in absence of the external magnetic
field and for the largest system size ($L=30$) simulated (Fig. \ref{f4}).
This observation thus does not provide any conclusive evidence 
about the nature of the nematic-isotropic transition for the system of
biaxial molecules. This is true for the system sizes we have investigated
and it is likely that using systems of significantly larger size one could
observe the free energy barrier for $\lambda=0.2$ and its dependence 
on the magnetic field.  
 
\subsection{Finite size scaling and the shift in NI transition temperature
with increasing magnetic field}
\begin{figure}[tbh]
\begin{center}
\resizebox{180mm}{!}{\rotatebox{-0}{\includegraphics[scale=1.0]{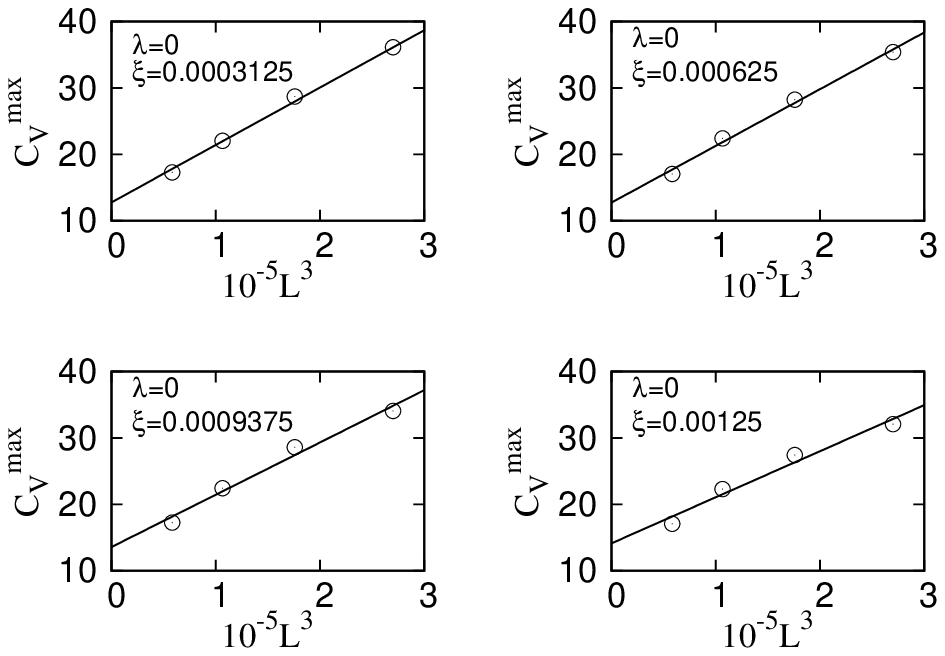}}}
\end{center}
\caption{Finite-size scaling behaviours of the peak height of $C_v$
for the uniaxial molecules for four different values of the field strength 
parameter $\xi=0.0003125-0.00125$. 
The estimated errors are of the order of the size of the symbols.
}\label{f5} 
\end{figure}

\begin{figure}[tbh]
\begin{center}
\resizebox{180mm}{!}{\rotatebox{-0}{\includegraphics[scale=1.0]{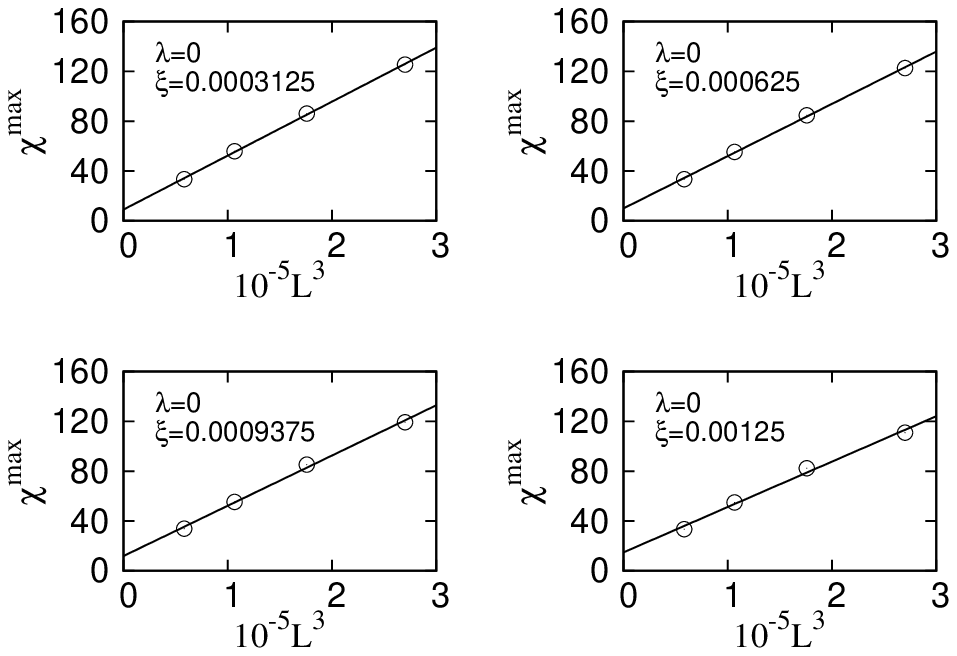}}}
\end{center}
\caption{Finite-size scaling behaviours of the peak height of $\chi$
for the uniaxial molecules for four different values of the field strength 
parameter $\xi=0.0003125-0.00125$.
The estimated errors are of the order of the size of the symbols.
}\label{f6} 
\end{figure}

For finite systems the peak height of $C_v$ increases with increasing 
system size and the scaling relation for $C_v^{max}$ 
in a first-order phase transition obeys $C_v^{max}\sim L^d$ \cite{lee1,lee2}.
We see that this scaling relation holds good for  
the uniaxial system (Fig. \ref{f5}). The data fit well 
for all values of $\xi$ used in our simulations.  
The expected linear fits of $\chi^{max}$ are presented in Fig. \ref{f6} for 
the uniaxial molecules. 
We have observed that with increasing $\xi$ the heights of 
the maxima $C_v^{max}(L)$ and $\chi^{max}(L)$ are suppressed 
particularly for the higher lattice sizes $L=26$ and $30$.
This observation shows that the presence of magnetic field weakens the 
first orderedness of the $N-I$ transition.

\begin{figure}[tbh]
\begin{center}
\resizebox{180mm}{!}{\rotatebox{-0}{\includegraphics[scale=1.0]{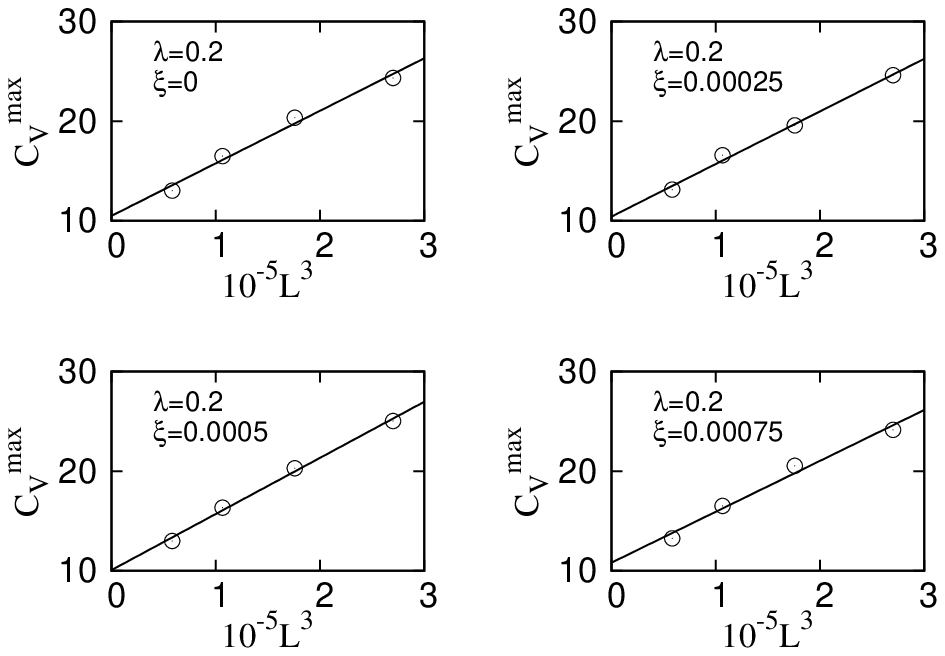}}}
\end{center}
\caption{Finite-size scaling behaviours of the peak height of $C_v$
for the biaxial molecules for four different values of the field strength 
parameter $\xi=0-0.00075$.
The estimated errors are of the order of the size of the symbols.
}\label{f7} 
\end{figure}

\begin{figure}[tbh]
\begin{center}
\resizebox{180mm}{!}{\rotatebox{-0}{\includegraphics[scale=1.0]{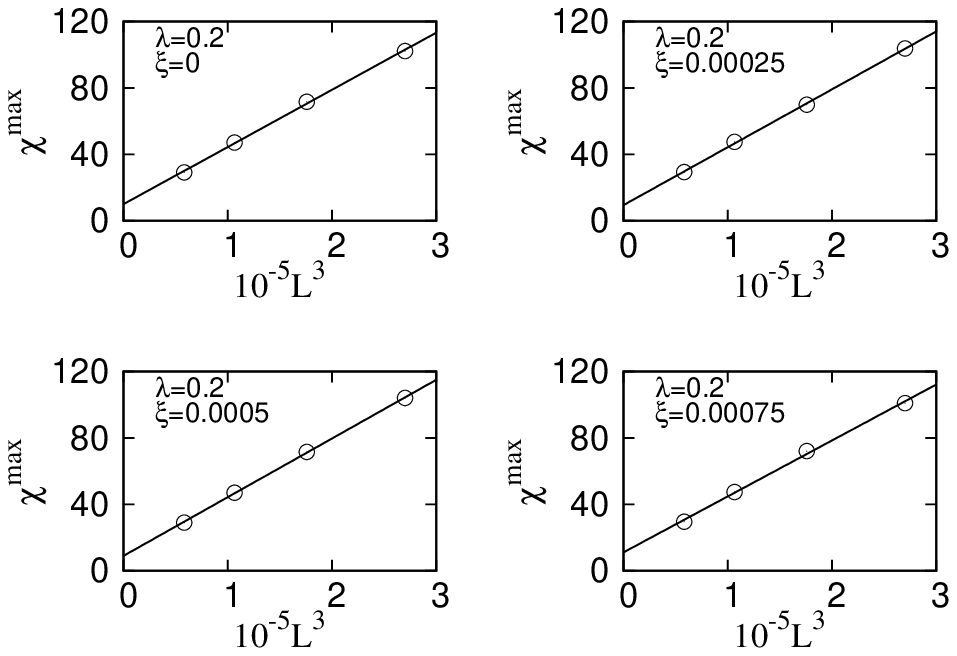}}}
\end{center}
\caption{Finite-size scaling behaviours of the peak height of $\chi$
for the biaxial molecules for four different values of the field strength 
parameter $\xi=0-0.00075$.
The estimated errors are of the order of the size of the symbols.
}\label{f8} 
\end{figure}

\begin{figure}[tbh]
\begin{center}
\resizebox{180mm}{!}{\rotatebox{-0}{\includegraphics[scale=1.0]{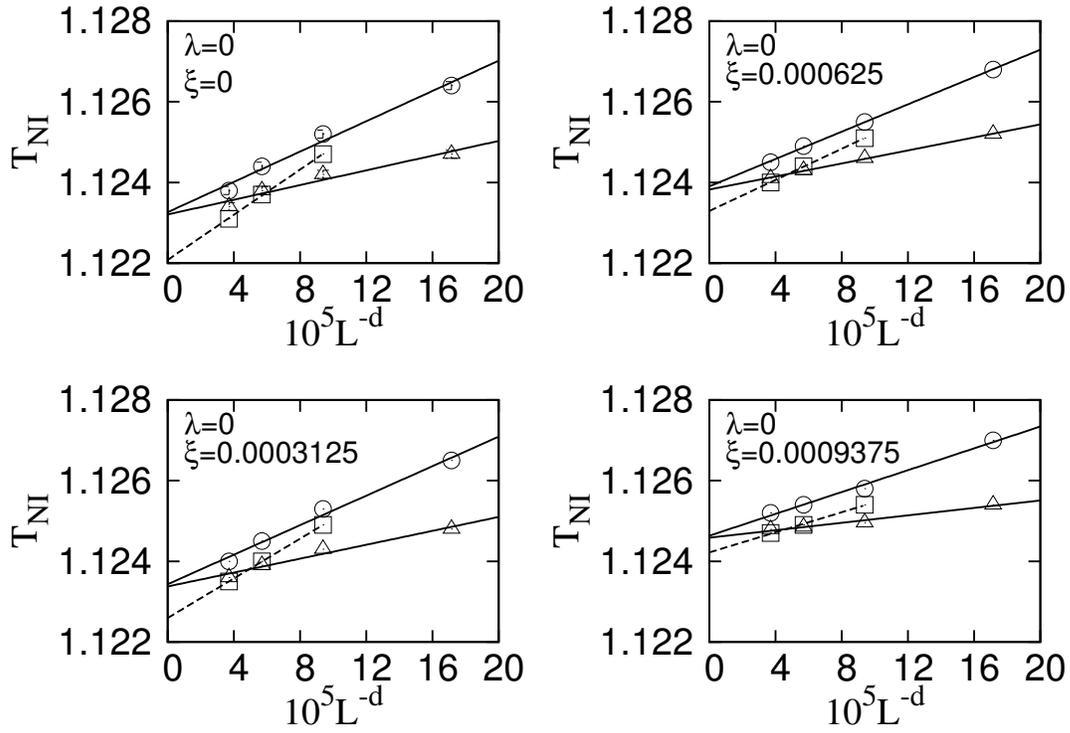}}}
\end{center}
\caption{Finite-size scaling behaviour shown for two different measures 
of $T_{NI}(L)$ for the uniaxial system at four different values of $\xi$
(the data points represented by $\vartriangle$ and $\circ$ are obtained from the peaks of $C_v$ and $\chi$ respectively). The estimates of  $T_-(L)$
are also shown ($\square$).
Extrapolations to the thermodynamic limits are denoted 
by solid and dashed lines, respectively.}\label{f9} 
\end{figure}

\begin{figure}[tbh]
\begin{center}
\resizebox{180mm}{!}{\rotatebox{-0}{\includegraphics[scale=1.0]{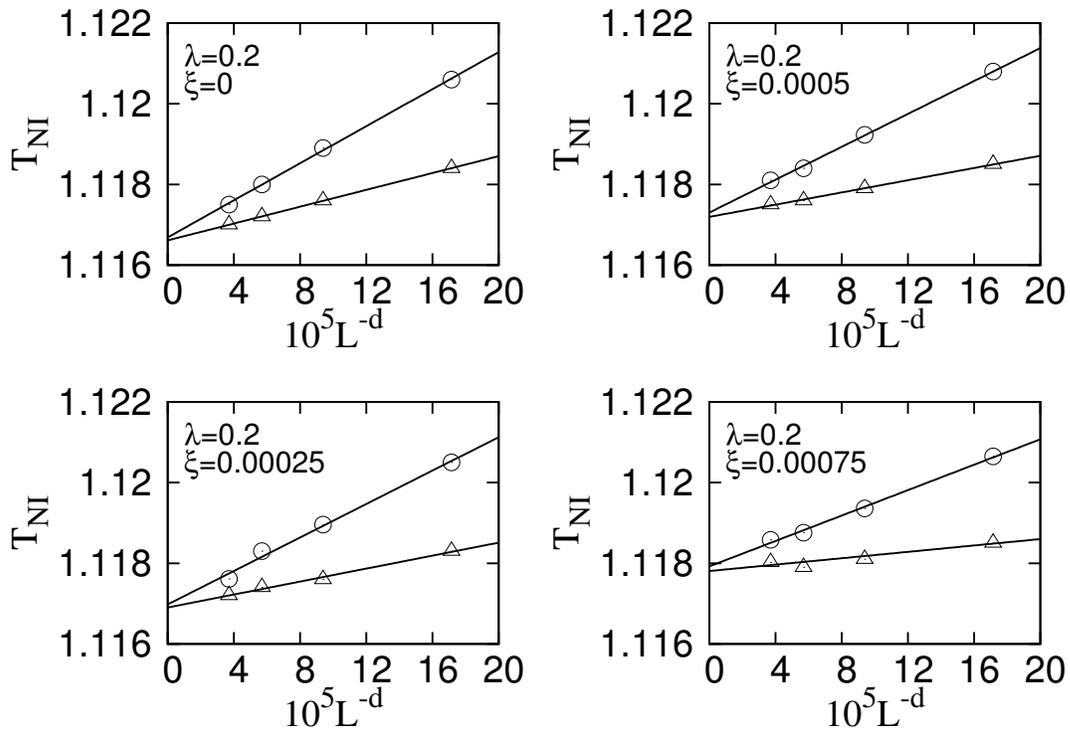}}}
\end{center}
\caption{Finite-size scaling behaviour shown for two different measures 
of $T_{NI}(L)$ for the biaxial system at four different values of $\xi$
(the data points represented by $\vartriangle$ and $\circ$ are obtained from the peaks of $C_v$ and $\chi$ respectively). 
Extrapolations to the thermodynamic limits are denoted 
by solid lines.}\label{f10} 
\end{figure}

\begin{figure}[tbh]
\begin{center}
\resizebox{115mm}{!}{\rotatebox{-90}{\includegraphics[scale=1.0]{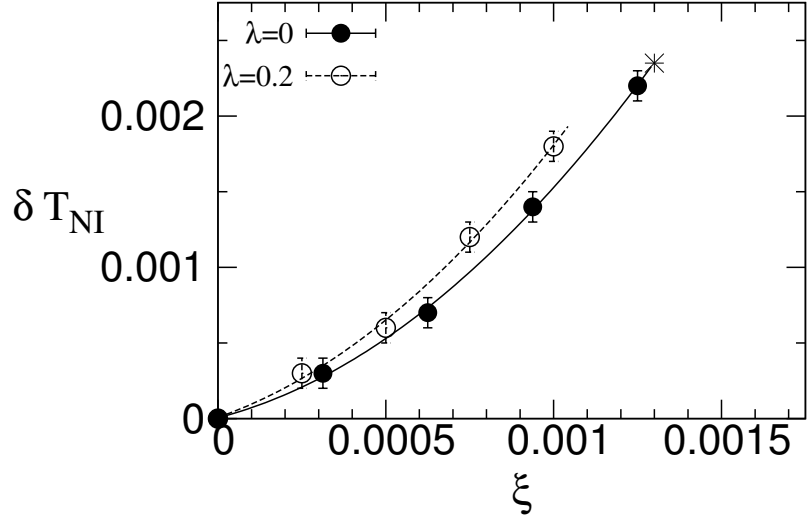}}}
\end{center}
\caption{Increase in the NI phase transition temperature $\delta T$ 
(over zero-field) vs $\xi$ for both the uniaxial ($\lambda =0$) 
and biaxial ($\lambda=0.2$) models. The extrapolation of the fitted
curve for $\lambda=0$ gives $\delta T_{NI}=0.00235$ for the critical parameter
$\xi_C=0.00130$ and the symbol ($\star$) represents the critical end point.}\label{f11}
\end{figure}

In case of biaxial system the straight line fits (Figs. \ref{f7} and \ref{f8}) to the data
of the peak heights of $C_v$ and $\chi$ for $\xi=0, 0.00025, 0.0005$
and $0.00075$ show that the $N-I$ transition 
obeys the first-order scaling laws for this system too.
Determining the finite-size transition temperature ($T_{NI}(L)$)
from the location of the maximum of the specific heat and  
the suceptibility curves
we have used the scaling relation $T_{NI}(L)-T_{NI}\sim L^{-d}$ and
performing a linear extrapolation to the thermodynamic limit 
($L\rightarrow\infty$)   
the transition temperature $T_{NI}$ in the thermodynamic limit was estimated.
In Fig. \ref{f9} the linear fits of the data for 
$T_{NI}(L)$ obtained from 
both $C_v$ and $\chi$ plots for the uniaxial molecules are shown
and in Fig. \ref{f10} the same are plotted 
for the biaxial molecules.
The estimated transition temperatures at the thermodynamic limit are listed 
in Table-1. We find the expected increase in $T_{NI}$ with increase 
in $\xi$ for both $\lambda=0$ and $\lambda=0.2$. 
The shift $\delta T_{NI}$ in the transition temperature is plotted 
against $\xi$ ($\xi \propto B^2$) in  Fig. \ref{f11}. 
We obtain good quadratic fits to the data in both cases.
The quadratic variations of $\delta T_{NI}$ with $\xi$ do not 
follow the prediction of the 
Landau de Gennes model for the isotropic-nematic transition 
which predicts a linear variation
of $\delta T_{NI}$ with $B^2$. The quadratic functions used for the fits
are $f(\xi)=5.7 \times 10^{-6}+0.57\xi+950.86\xi^2$ for $\lambda=0$ and
$g(\xi)=8.5 \times 10^{-6}+0.77\xi+1028.57\xi^2$ for $\lambda=0.2$.
Therefore, the rise of $T_{NI}$ over its zero-field value is steeper 
for the biaxial molecules than 
that for the uniaxial molecules. 
For low magnetic fields the slope of the fit for $\lambda=0.2$ is $35\%$ higher than 
that for $\lambda=0$.
The results suggest
that the external magnetic field necessary for inducing 
the $I (pN)-N$ phase transition is lower 
for the biaxial system than the uniaxial system
(where initial temperatures for both systems are assumed to be equally 
higher than their respective zero-field transition temperatures, $T_{NI}(0)$). 
The mean field
study of Wojtowicz and Sheng \cite{woj} for the uniaxial model shows that 
the slope of a linear fit is lower by $20\%$ in comparison to 
the slope extracted from Fig.$3$ of Ref.\cite{rem} in which the model
under investigation was biaxial.  
Our observation regarding the shift in $T_{NI}$ with increase in magnetic field 
is therefore consistent with the mean field results. 

The finite size stability limit, $T_-(L)$ of the isotropic phase 
for uniaxial molecules is estimated as the temperature where the second, 
local minimum (at higher energy) of $A$ just vanishes as $T$ is gradually 
lowered below $T_{NI}$. From extrapolation to the thermodynamic limit 
we estimate $T_-$ for different field strengths as indicated 
in Fig. \ref{f9} by dashed lines. 
%

\subsection{Estimate of the critical field}
\begin{figure}[tbh]
\begin{center}
\resizebox{115mm}{!}{\rotatebox{0}{\includegraphics[scale=1.0]{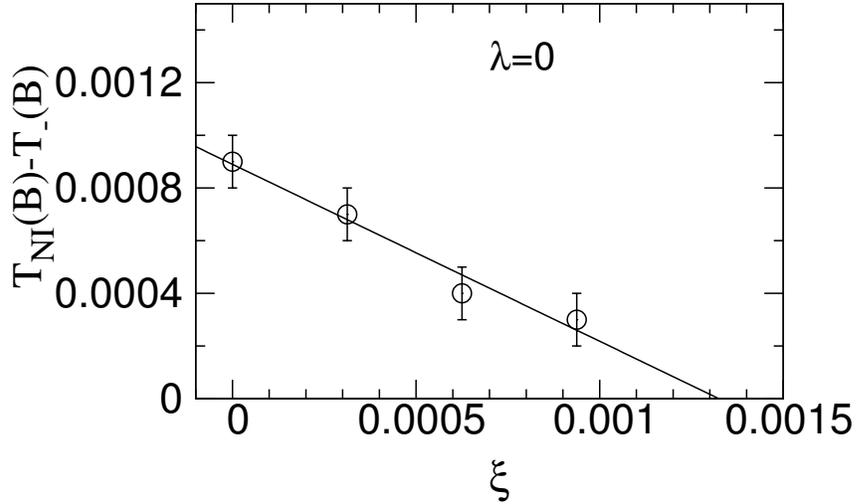}}}
\end{center}
\caption{Plot of $T_{NI}(B)-T_-(B)$ vs $\xi$. 
The solid line is the best linear fit. The estimated value of $\xi_C$
is $0.00130$.}\label{f12} 
\end{figure}

An important observation is that the width of the stability limit 
of the $I$ phase, $T_{NI}(B)-T_-(B)$ decreases with increase
in magnetic field (Fig. \ref{f12}). 
From the fit to the data (as shown in Fig. \ref{f12}) we find 
that the temperature difference $T_{NI}(B)-T_-(B)$ varies linearly with
the field parameter $\xi$ (i.e., quadratic in $B$).
With increasing magnetic field the weak first order $N-I$ transition of 
the uniaxial model becomes weaker and $T_-(B)$ gradually becomes 
closer to $T_{NI}(B)$ and at the critical end point they merge into 
a single point. It is, therefore, possible to obtain an estimate 
of the critical magnetic field by performing a linear extrapolation 
(Fig. \ref{f12}) up to the field strength ($\xi_C$) at which 
$T_{NI}(B)-T_-(B)$ becomes zero. For the uniaxial case we get 
$\xi_C=0.00130$. 
In order to get an estimate of $\xi$ in real units we may express the 
magnetic field as $B=(3\mu_0\epsilon\xi/\Delta\kappa)^{1/2}$.
The energy unit $\epsilon$ can be estimated by using the experimental 
and simulated nematic-isotropic transition temperatures, i.e. 
$\epsilon=k_BT_{K}(B=0)/T_{NI}(B=0)$. For a common nematic, say  
MBBA, the $N-I$ transition temperature $T_{K}(B=0)=320K$ 
and the anisotropy of the molecular magnetic polarizability
$\Delta \kappa \approx 0.16\times 10^{-32}$~$m^3$ \cite{deju}.
Using these values and the simulated value of $T_{NI}(B=0)=1.1232$ 
for $\lambda=0$ we can obtain $B\approx3040\sqrt{\xi}$~$T$.
Therefore, the critical magnetic field $B_C$ for common nematics 
corresponding to our estimate,
$\xi_C=0.00130$, is $\sim 110~T$.

\begin{figure}[tbh]
\begin{center}
\resizebox{110mm}{!}{\rotatebox{0}{\includegraphics[scale=1.0]{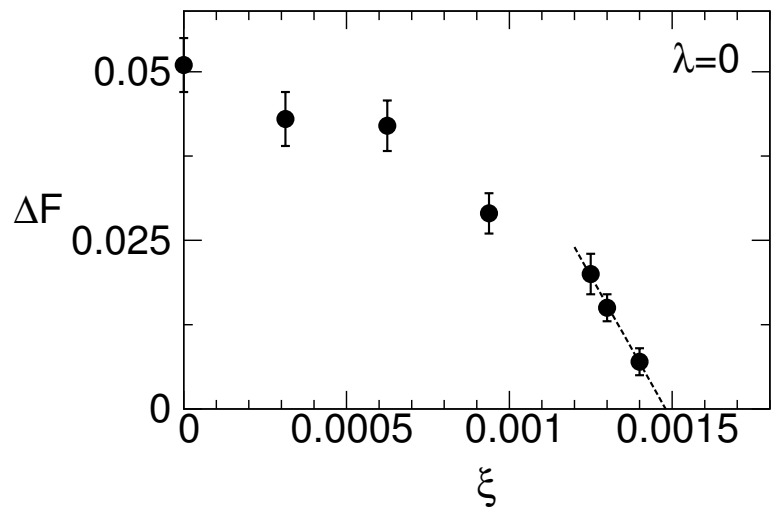}}}
\end{center}
\caption{Free energy barrier height $\Delta F(L)$ vs $\xi$
for the uniaxial model and for the lattice size $L=30$.
A linear extrapolation of three nearest points is used to estimate 
$\xi_C$ which is $0.00148$.}\label{f13}
\end{figure}

We can also obtain a finite size estimate of $\xi_C(L)$ from   
the field dependence of $\Delta F$ \cite{saito} as shown in Fig. \ref{f13}.
We observe that as the field strength parameter $\xi$ is 
increased beyond $0.00125$ the free energy
barrier decreases much faster.
For the uniaxial molecules we may estimate the value of $\xi_{c}$ 
by performing a linear 
extrapolation using the results of the barrier height ($\Delta F$)
for three higher values of $\xi$ i.e. $0.00125, 0.0013$ 
and $0.0014$ (we have extended our simulations to $\xi=0.0013$ and 
$0.0014$ for the largest lattice size $L=30$). 
Our estimate of the finite size critical field parameter is $\sim 0.00148$
which results in a value of the critical field $B_C$ 
for the common nematogens to be
$\sim 120~T$. This is larger than the estimate in thermodynamic limit 
given in the previous paragraph.  
Our estimate of the critical magnetic field is  
of the same order of magnitude
($130~T$) mentioned in \cite{lel1} obtained from the phenomenological
Landau-de Gennes theory and using the experimental values of the $f$ 
parameters for $5CB$. But the critical magnetic field predicted by 
the molecular theory of Maier-Saupe \cite{woj} is much higher 
($1000~T$) than our estimate.  

The corresponding critical  
temperature can be estimated from the extrapolation of the fitted curve for 
$\lambda=0$ in Fig.\ref{f11} which gives $T_C=1.12555$, $\delta T_C$ 
being $0.00235$.

\section{CONCLUSIONS}
We have determined the shift in the $N-I$ transition 
temperature with increasing external magnetic field
in the thermodynamic limit for two nematic systems composed of uniaxial 
and biaxial molecules. These results represent the first 
numerical evidence of the first order nature of nematic-isotropic
transition in an external magnetic field the presence of which makes
the first order character weaker. The first order nature of the transition
is established from the finite-size scaling behaviour of the transition 
temperature $T_{NI}(L)$ obtained from the specific heat 
and susceptibility data. 

The quadratic dependence of $\delta T_{NI}$ with $\xi$ where $\xi \propto B^2$, 
for both cases $\lambda=0$ and $\lambda=0.2$,
does not agree with the theoretical prediction of LDG 
model which predicts this to be linear for low values of 
the external magnetic field.  
The steeper nature of the quadratic fit for the biaxial molecules in 
comparison to that of the uniaxial molecules shows 
that a lower magnetic field strength is capable of inducing 
the isotropic-nematic transition for the biaxial case.

The free energy like quantity generated by the  
reweighting technique gives detailed insight of the equilibrium  
properties of the models in presence of an 
external magnetic field.
In particular, for the uniaxial case, the effect of an external magnetic field 
on the free energy barrier is clearly observed.     
Our results show that the width of the stability limit of the 
isotropic phase below $T_{NI}(B)$ decreases quadratically with the 
external field.  
We estimate the critical magnetic field strength for a system of uniaxial 
mesogenic molecules ($\sim 110$~$T$).  
For the biaxial system we have not found any double well 
structure in the free energy. 
The work further suggests that the first
order nature of the nematic-isotropic transition of a system of
biaxial molecules is considerably weaker than that
shown by the uniaxial LL model.
The presence of an external magnetic field further weakens the first orderedness
of the NI transition.
To study the effect of magnetic field on the nematics composed of 
biaxial molecules systems of much bigger sizes need to be simulated.

\section{ACKNOWLEDGMENT}
NG acknowledges the award of a teacher fellowship
under Faculty Development Programme of UGC, India, and KM is grateful
to the UGC for the award of a minor research project.

\begin{table}[h]
\caption{NI transition temperatures (at thermodynamic limit) for 
different values of the external field strength parameter $\xi$ for the uniaxial and the biaxial systems. Estimates of the super cooling limits $T_-$
are also listed for the uniaxial system. 
The estimated (jackknife) error in each 
temperature is $\pm 0.0001$.}
\begin{center}
\begin{tabular}{c c c c c}
\hline
$\lambda$ & $\xi$ & $T_{NI}$(from~$C_V{\it vs}T$) & $T_{NI}$(from~$\chi{\it vs}T$) & $T_-$\\
\hline
 &0 &1.1232 &1.1232 & 1.1221 \\
 &0.0003125 &1.1234 & 1.1234& 1.1226 \\
0 &0.0006250 &1.1238 &1.1239 & 1.1233\\
 &0.0009375 &1.1245 &1.1245 & 1.1242\\
 &0.0012500 &1.1254 &1.1255 &\\
\hline
 &0 &1.1166 &1.1167  \\
 &0.00025 &1.1169 &1.1170  \\
0.2 &0.00050 &1.1172 &1.1173  \\
 &0.00075 &1.1178 &1.1179  \\
 &0.00100 &1.1184 &1.1185  \\
\hline
\end{tabular}
\end{center}
\end{table}


\begin{thebibliography}{99}
\bibitem{fre} M. J. Freiser and R. J. Joenk, Phys. Letters {\bf24A}, 683 (1967).
\bibitem{han} J. Hanus, Phys. Rev. {\bf178}, 420 (1969).
\bibitem{fan} C. P. Fan and M. J. Stephen, Phys. Rev. Lett. {\bf25}, 500 (1970).
\bibitem{woj} P. J. Wojtowicz and P. Sheng, Phys. Letters {\bf48A}, 235 (1974).
\bibitem{she} J. Shen and C. W. Woo, Phys. Rev. A {\bf24}, 493 (1981).
\bibitem{hor} R. M. Hornreich, Phys. Letters {\bf109A}, 232 (1985).
\bibitem{rem} D. K. Remler and A. D. J. Haymet, J. Phys. Chem. {\bf90}, 5426 (1986).
\bibitem{hel} W. Helfrich, Phys. Rev. Lett. {\bf24}, 201 (1970).
\bibitem{nic} A. J. Nicastro and P. H. Keyes, Phys. Rev. A {\bf30}, 3156 (1981).
\bibitem{lel1} I. Lelidis and G. Durand, Phys. Rev. E {\bf48}, 3822 (1993).
\bibitem{lel2} I. Lelidis, M. Nobili and G. Durand, Phys. Rev. E {\bf48}, 3818 (1993).
\bibitem{rosen} C. Rosenblatt, Phys. Rev. A {\bf24}, 2236 (1981).
\bibitem{osta} T. Ostapenko, D. B. Wiant, S. N. Sprunt, A. Jakli, and J. T. Gleeson, Phys. Rev. Lett. {\bf101}, 247801 (2008).
\bibitem{degen} P. G. de Gennes and J. Prost, $\textit{The Physics of Liquid Crystals}$, (Oxford Science, Oxford, 1993).
\bibitem{stanley} H. E. Stanley, $\textit{Introduction to Phase Transition and Critical Phenomena}$, (Oxford University Press, Oxford, 1971).
\bibitem{dhara} S. Dhara and N. V. Madhusudana, Eur. Phys. J. E {\bf22}, 139 (2007).
\bibitem{shen} J. Shen and C. Woo, Phys. Rev. A {\bf24}, 493 (1981).
\bibitem{madsen} L. A. Madsen, T. J. Dingemans, M. Nakata, and E. T. Samulski, Phys. Rev. Lett. {\bf92}, 145505 (2004).
\bibitem{achar} B. R. Acharya, A. Primak, and S. Kumar, Phys. Rev. Lett. {\bf92}, 145506 (2004).
\bibitem{link} D. R. Link, G. Natale, R. Shao, J. E. Maclennan, N. A. Clark, E. Korblova, and D. M. Walba, Science {\bf278}, 1924 (1997).
\bibitem{wiant} D. Wiant, S. Stojadinovic, S. Sharma, K. Fodor-Csorba, A. Jakli, J. T. Gleeson, and S. Sprunt, Phys. Rev. E {\bf73}, 030703(R) (2006).
\bibitem{ferr} A. M. Ferrenberg and R. H. Swendsen, Phys. Rev. Lett. {\bf61}, 2635 (1988); {\bf63}, 1195 (1989).
\bibitem{lee1} J. Lee and J. M. Kosterlitz, Phys. Rev. Lett. {\bf65}, 137 (1990).
\bibitem{lee2} J. Lee and J. M. Kosterlitz, Phys. Rev. B {\bf43}, 3265 (1991).
\bibitem{saito} H. Saito, S. Ejiri, S. Aoki, T. Hatsuda, K. Kanaya, Y. Maezawa,
 H. Ohno, and T. Umeda, Phys. Rev. D {\bf84}, 054502 (2011).
\bibitem{luck1} G. R. Luckhurst, P. Simpson and C. Zannoni, Chem. Phys. Lett. {\bf78}, 429 (1981); G. R. Luckhurst and G. Saielli, J. Chem. Phys., {\bf112}, 4342 (2000).
\bibitem{berar} R. Berardi, S. Orlandi and C. Zannoni, Mol. Cryst. Liq. Cryst.,
P. L. Nordio Memorial issue, 2002.
\bibitem{buck} A. D. Buckingham, in $\textit{Intermolecular Forces}$, edited 
by J. O. Hirschfelder (Wiley, London, 1967), Chap. 2 [Adv. Chem. Phys. {\bf12},
 107 (1967)].
\bibitem{stone} A. J. Stone, $\textit{The Theory of Intermolecular Forces}$,  
(Oxford University Press, Oxford, UK, 1997).
\bibitem{luck} G. R. Luckhurst and S. Romano, Mol. Phys. {\bf40}, 129 (1980).
\bibitem{bis} F. Biscarini, C. Chiccoli, P. Pasini, F. Semeria, and C. Zannoni, Phys. Rev. Lett. {\bf75}, 1803 (1995).
\bibitem{rose} M. E. Rose, $\textit{Elementary Theory of Angular Momentum}$,
(Wiley, New York, 1957).
\bibitem{leb} P. A. Lebwohl and G. Lasher, Phys. Rev. A {\bf6}, 426 (1972).
\bibitem{zhang} Z. Zhang, O. G. Mouritsen, and M. J. Zuckermann, Phys. Rev. Lett. {\bf69}, 2803 (1992).
\bibitem{fab} U. Fabbri and C. Zannoni, Mol. Phys. {\bf58}, 763 (1986).
\bibitem{ghoshal} N. Ghoshal, K. Mukhopadhyay and S. K. Roy, Liq. Cryst. {\bf39}, 1381 (2012); arXiv:1205.6639v1 [cond-mat.soft] (2012).
\bibitem{bar} J. A. Barker and R. O. Watts, Chem. Phys. Lett. {\bf3}, 144 (1969).
\bibitem{new} M. E. J. Newman and G. T. Barkema, $\textit{Monte Carlo Methods in Statistical Physics}$,
(Clarendon press, Oxford, 1999).
\bibitem{vie} J. Vieillard-Baron, J. Chem. Phys. {\bf56}, 4729 (1972).
\bibitem{camp} P. J. Camp and M. P. Allen, J. Chem. Phys. {\bf106}, 6681 (1997).
\bibitem{deju} W.~H.~de Jeu, Physical properties of liquid crystalline
materials, Gordon and Breach Science Publishers, London, 1979.
\end{thebibliography}
\end{document}